Quantum entanglement for two electrons in the excited states of helium-like systems

Yen-Chang Lin[1,2,*] and Yew Kam Ho[1,**]

[1] Institute of Atomic and Molecular Sciences, Academia Sinica, Taipei, Taiwan
[2] Graduate Institute of Applied Science and Engineering, Fu-Jen Catholic University, Taiwan

**Abstract**

Quantum entanglement for the two electrons in excited states of the helium-like atom/ions is investigated using two-electron wave functions constructed by the *B*-Spline basis. As a measure of spatial (electron-electron orbital) entanglement, the von Neumann entropy and linear entropy of the reduced density matrix are calculated for the $1s2s$ $^{1,3}S$ excited states for systems with some selected $Z$ values from $Z=2$ to $Z=100$. Results for the helium atom are compared with other available calculations. We have also investigated the entropies for these excited states when the nucleus charge is reduced from $Z=2$ continuously to $Z=1$. At such a critical charge, all the singly-excited states of this system become unbound, and the linear entropies and the von Neumann entropies for the excited states are approaching 1/2 and 1, respectively, the limits for the entropies when one electron is bound to the nucleus, and the other being free.

# I. INTRODUCTION

There has been considerable interest in the investigations of quantum entanglement for bi-particle atomic systems. Understanding of entangled systems is important in other research areas such as quantum information [1], quantum computation [2] and quantum cryptography [3]. The investigations of quantum entanglement include the works on some model atoms like the Moshinsky atom [4-8], the Crandall atom [9] and the Hooke atom [9-12], and the works on artificial atoms like quantum dots [13-17]. Coe and D'Amico calculated the linear entropy for the ground state of the natural helium atom with wave functions constructed by using the products of hydrogenic wave functions, as well as using the density functional theory [10]. Manzano *et al*. [9] and Dehesa *et al.* [18] investigated the entanglement of the helium ground and excited states using Kinoshita-type wave functions. Benenti *et al* [19] calculated the linear entropy and von Neumann entropy for the helium atom using configuration interaction (CI) basis wave functions constructed with Slater-type orbitals (STO). Huang *et al* [20] using the Gaussian basis sets calculated the von Neumann entropy for the helium atom and the $H_2$ molecule. Lin *et al* [21] calculated the linear entropy for the ground and singlet-spin excited states for the helium-like ions by employing configuration interaction wave functions constructed with *B*-Spline basis. The linear entropy for the ground states of three-body atomic systems such as the hydrogen negative ion (also called hydride), $H^-$, the positronium negative ion and the helium atom were calculated by employing highly-correlated Hylleraas wave functions [22]. Hylleraas-type wave functions were also used in recent works to calculate entropies for helium atom [23] and helium-like systems [24, 25]. Hofer used Gaussian wave functions to examine basis set convergence for electron-electron entanglement in helium-like systems [26]. Osenda and Serra have investigated the effect of critical charge on the ground and excited states for the *S*-wave model of helium atom [27, 28]. Tichy *et al* [29] has reviewed the recent developments on entanglement for atomic and molecular systems. In related developments, investigations on quantum entanglement of two electrons in coupled quantum dots were reported in the literature [30, 31]. All the above mentioned works on bipartite atomic systems have played an important role in discussions of a current issue of how to quantify entanglement entropies for indistinguishable particles such as identical fermions or identical bosons [32-40].

In the present work, we report calculations of von Neumann entropy ($S_{vN}$) for the ground state of the helium atom. For the excited singlet-spin $1s2s$ $^1S$ and triplet-spin $1s2s$ $^3S$ states, we calculate the von Neumann entropy ($S_{vN}$) and linear entropy ($S_L$) for the helium-like systems with some selected values of *Z* up to *Z*=100. Furthermore, we report an investigation of critical charge effect on $S_L$ and $S_{vN}$ for the excited $1s2s$ $^{1,3}S$ states of helium when the nuclear charge is reduced from *Z*=2 continuously to *Z*=1. For the helium atom (*Z*=2), there are infinite number of bound states in the system due to the Coulomb nature of the potential. When *Z* is reduced continuously to the critical charge *Z*=1, the system now consists of a proton and two electrons and has only one bound state, the ground state (denoted as $1s^2$ $^1S^e$) of the $H^-$ ion. All the excited states in the cases for *Z* > 1 are now becoming unbound when Z=1. Here, we report an investigation on the behaviors of entanglement entropies for the abovementioned excited states when Z is approaching the critical charge. Atomic units (a. u.) are used throughout the present work.

## II. THEORETICAL METHOD

In this work, we use von Neumann and linear entropies to quantify quantum entanglement for the ground and excited stats of the helium atom. The von Neumann entropy, denoted as $S_{vN,}$, has the form

$$S_{vN} = -\text{Tr}(\rho \ln \rho), \tag{1}$$

with maximum entropy equals to ln(2) in a two levels or a two particles system. It has also been expressed in literature [19, 20, 23, 25, 26], in a form with base $\log_2$, as

$$S_{vN} = -c \, \text{Tr}(\rho \ln \rho) = -\text{Tr}(\rho \log_2 \rho), \tag{2}$$

where $\rho$ is density matrix and $c = 1/\ln(2)$. The linear entropy, denoted as $S_L$ and is defined in a usual form in the literature, as

$$S_L = 1 - \text{Tr}\rho^2 . \tag{3}$$

$S_L$ can be considered as an approximation of the von Neumann entropy (dropping the constant $c$ for now) by taking the leading term in the expansion of $\ln \rho$, and the linear entropy becomes

$$S_L = -\text{Tr}(\rho(\rho-1)) = \text{Tr}(\rho - \rho^2) = Tr(\rho) - Tr(\rho^2) = 1 - \text{Tr}(\rho^2). \tag{4}$$

Here, $\text{Tr}(\rho^2)$ is referred to as the purity of the state and in general $\rho$ is for the total system, but entanglement reflects the relationship between different parts in the system. Now in Eq. (2) we replace $\rho$ by $\rho_{red}$, where $\rho_{red}$ is the one-particle reduced density matrix. For two-component quantum systems, e.g., two-electron atoms, the one-particle reduced density matrix is obtained by tracing the two-particle density matrix over all the degrees of freedom for one of the two particles. We have

$$\rho_{red}(\vec{r}_1, \vec{r}_2) = \int \psi^*(\vec{r}_1, \vec{r}_3) \psi(\vec{r}_2, \vec{r}_3) d\vec{r}_3 \tag{5}$$

and

$$\rho_{red}^2(\vec{r}_1, \vec{r}_2) = \int \rho_{red}(\vec{r}_1, \vec{r}_3) \rho_{red}(\vec{r}_2, \vec{r}_3) d\vec{r}_3 \tag{6}$$

with

$$\text{Tr}\rho_{red}^2 = \int \rho_{red}^2(\vec{r}, \vec{r}) d\vec{r} . \tag{7}$$

The Hamiltonian for the helium-like system is

$$H(r_1,r_2) = -\frac{1}{2}\nabla_1^2 - \frac{1}{2}\nabla_2^2 - \frac{Z}{r_1} - \frac{Z}{r_2} + \frac{1}{r_{12}}, \tag{8}$$

where the index 1 and 2 are for the two electrons, respectively, and $Z=2$ for the helium atom. For two-electron atoms, the basis function $\psi_{nl,n'l'}^{\Lambda}(r_1,r_2)$ with $\Lambda$ standing for a set of quantum number ($S$, $L$, $M_S$, $M_L$), can be constructed through the expansion of two-particle Slater-determinant wave functions [41], i.e.,

$$\psi_{nl,n'l'}^{\Lambda}(r_1,r_2) = \sum_{mm',m_s m'_s} (-1)^{l'-l} [(2S+1)(2L+1)]^{1/2} \begin{pmatrix} l & l' & L \\ m & m' & -M_L \end{pmatrix} \begin{pmatrix} \frac{1}{2} & \frac{1}{2} & S \\ m_s & m'_s & -M_S \end{pmatrix} \phi_{nl,n'l'}^{mm_s,m'm'_s}(r_1,r_2) \tag{9}$$

where the Slater-determinant wave function $\phi_{nl,n'l'}^{mm_s,m'm'_s}(r_1,r_2)$ is given by the one-particle orbital $u_{nl}^{mm_s}(r)$ including spin, i.e.,

$$\phi_{nl,n'l'}^{mm_s,m'm'_s}(r_1,r_2) = \frac{1}{\sqrt{2}} \left[ u_{nl}^{mm_s}(r_1) u_{n'l'}^{m'm'_s}(r_2) - u_{nl}^{mm_s}(r_2) u_{n'l'}^{m'm'_s}(r_1) \right] \tag{10}$$

The configuration interaction method chooses the basis for non-zero wave function of the Wigner 3-j symbol, $\begin{pmatrix} l & l' & L \\ m & m' & -M_L \end{pmatrix}$ and $\begin{pmatrix} 1/2 & 1/2 & S \\ m_s & m'_s & -M_S \end{pmatrix} \neq 0$. And the parity of wave function is non-zero term, $\phi(r_1,r_2) - P_{12}\phi(r_1,r_2) \neq 0$, $P_{12}$ is parity operator. The radial function $\chi_{nl}(r)$ of the one-particle orbital is expanded in terms of the B-Spline basis functions $B_{i,k}(r)$, i.e.,

$$\chi_{nl}(r) = \sum_{i=1}^{N} C_i B_{i,k}(r) \tag{11}$$

Along the $r$ axis with end points $r=0$ and $r=R$, we select a knot sequence $\{t_i\}$ $(i=1,2,3,\cdots,N+k)$, with $r_{min} \leq t_i \leq r_{max}$ and $t_i \leq t_{i+1}$. In the present work $t_i = r_{min}$ ($i=1$ to k) and $t_i = r_{max}$ ($i=n+1$ to $n+k$). The B-Spline basis functions of order $k$ are defined on the knot sequence by the following recurrence relations [42]:

$$B_{i,1}(r) = \begin{cases} 1 & \text{for } t_i \leq r < t_{i+1} \\ 0 & \text{otherwise} \end{cases} \tag{12}$$

and

$$B_{i,k}(r) = \frac{r-t_i}{t_{i+k-1}-t_i} B_{i,k-1}(r) - \frac{t_{i+k}-r}{t_{i+k}-t_{i+1}} B_{i+1,k-1}(r). \tag{13}$$

The boundary conditions of the radial wave functions require $\phi(0) = \phi(R) = 0$, which leads to the vanished coefficients $C_1$ and $C_N$ because of the properties of the B-Spline at the two end points, i.e. $B_{1,k}(0)=1$ and $B_{N,k}(0)=1$. For more discussions about B-Spline basis functions, readers are referred to Ref. [42]. The computer code involving B-Spline basis functions for two-electron systems used in the present work has been developed in our group based on the formulism shown in Ref. [41]. The use of B-Spline basis for quantification of entanglement entropies for natural atoms in the present work and in Ref. [21] is part of continuing effort by our group throughout recent years to apply such basis functions to study various aspects of bound and continuum states in atomic physics, including a study of strong electric-field effects on the ground state photoionization of helium atom [43]; determination of resonance energies and widths of doubly excited resonant states in divalent magnesium atom [44]; an investigation of spectral properties of helium atoms with screened Coulomb potentials [45]; a calculation of bound-state energies, oscillator strengths, and multipole polarizabilities for the hydrogen atom with exponential-cosine screened Coulomb potentials [46]; and the recent evaluation of one- and two-photon ionization cross sections of hydrogen atom embedded in Debye plasmas [47].

In the present work we report calculations of $S_L$ and $S_{vN}$ for the ground state and the $1s2s$ $^{1,3}S$ excited states of the helium atom. The method of configuration interaction (CI) is used to construct the two-electron wave functions for the ground and excited states of helium atom. For the most part, such two-electron wave functions are obtained with one-electron basis functions including s, p, d, f, g, and h orbitals (the maximum l value is $l_{max}=5$) with the principal quantum number n, denoted as $n_{max}$, up to 40. The eigenvalues and eigenvectors for a total of about 4000 $^1S$ configurations are obtained by diagonalization of the Hamiltonian matrix. For some selected cases, like the case for helium, we have extended the calculation to include orbitals up to $l_{max}=6$ or 7 to test convergence of our results. We also point out that the present work is an extension of our recent paper [21] in which linear entropy for ground states in the helium-like systems were reported. More details of such CI-B-Spline basis wave functions can be found in our earlier publications [21, 43, 44, 45, 46, 47].

Finally, it should be mentioned that in the present work we only consider the spatial part of the two electron systems. For the contributions to the entanglement entropy from the spin part, readers are referred to the earlier publications [9-11, 17-18]. For self-contained manner, in Appendix A at the end of the text, we describe briefly the entanglement entropy attributed to the spin and spatial parts of the two-electron helium atom.

The reduce density matrix has eigenvalues $\lambda_i$ and eigenfunctions $\phi_i$ with

$$\int \rho_{red}(i,j)\phi(j)dj = \lambda_i \phi(i) . \tag{14}$$

Or, in matrix form

$$\sum_j \rho_{red,ij}\phi_j = \lambda_i\phi_i. \qquad (15)$$

For the wave functions that are constructed with products of separable one-electron basis, the reduced density matrix can be calculated with a series of matrix multiplications that involve the eigenvector of the state under investigation. In Appendix B at the end of the text, we show how reduced density matrixes for the singlet-spin and triplet-spin wave functions can be calculated. Once the elements for the reduced density matrix are determined, the eigenvalues $\lambda_i$ for such a matrix can be obtained, and the von Neumann entropy can then be determined with

$$S_{vN}(\rho_A) = -Tr(\rho_A \log_2 \rho_A) = -\sum_i \lambda_i \log_2 \lambda_i, \qquad (16)$$

and the linear entropy with

$$S_L(\rho_A) = 1 - Tr(\rho_A^2) = 1 - \sum_i \lambda_i^2. \qquad (17)$$

For a certain type of wave functions, the method using Eq. (17) for calculations of linear entropy represents an alternate, but simpler, way as compared to the scheme using Eqs. (3) and (7). In the present work, our results on von Neumann entropy and linear entropy are obtained by using Eqs. (16) and (17), respectively.

## III. CALCULATIONS AND RESULTS

Next we construct the wave functions for the ground state and the $1s2s$ $^{1,3}S$ excited states of the helium atom, and their energies are shown here in Table I. It is seen that the energies for our wave functions compare quite well with the most accurate non-relativistic energies in the literature [48]. In Tables 2 and 3 we show convergence tests on $S_{vN}$ and $S_L$, respectively, for the helium ground state. Calculations were done in personal computers (PC) with CPU made by Intel® with Core™ i7-3770 processor, having capacity of 16 GB RAM and CPU speed of 3.40GHz. For one-electron $l_{max} = 6$ and $n_{max} = 40$, the sizes of two-electron basis functions are 4935 and 4676 for $^1S$ and $^3S$ states, respectively. It takes a total time of about 67 minutes for $^1S$ states and 63 minutes for $^3S$ states of CPU time to complete the entropy calculation, including construction of wave function and determination of its energy, for one atomic state on double precision algorithm (64 bits long for word length). For one-electron orbitals used up to $l_{max} = 7$ and with $n_{max} = 35$, the size of the two-electron matrix is 5341 (for $^1S$ states) and 5054 (for $^3S$ states), and it takes about 98 minutes (for $^1S$) and 93 (for $^3S$) minutes, respectively, to complete calculation for the same state mentioned above. So for practical purposes, we have not extended our basis wave functions beyond these two limits for basis sizes. From the convergence tests shown in Tables 2 and 3 we conclude that our results for linear entropy and von Neumann entropy for the ground $1s^2$ $^1S$ state are $S_L = 0.015937 \pm 0.00004$ and for $S_{vN} =$

0.084998 ± 0.0001 respectively. Tables 4 and 5 show similar convergence tests on $S_L$ and $S_{vN}$, respectively, for the $1s2s$ $^1S$ state of He. From such convergence tests, we estimate the entropies for this state are $S_L$ = 0.488737 ± 0.00001 and $S_{vN}$ = 0.991917 ± 0.00001. Finally, in Tables 6 and 7, we show the convergence tests for the $1s2s$ $^3S$ state of He, leading to $S_L$ = 0.500376 ± 0.00001 and $S_{vN}$ = 1.005527 ± 0.00001. In Table 8, we show a comparison of $S_L$ and $S_{vN}$ with the earlier calculations. For the ground state, the present results calculated by using Eq. (17) are practically identical to our earlier results calculated by using Eqs. (3) and (7) (see Ref. [21]). The present results are obtained by diagonalizing the reduced density matrix, while the earlier results involved calculations of the square of the reduced density matrix. As for $S_{vN}$, we also compare our results with those in Ref. [19] in which configuration interaction basis wave functions were constructed by using the products of Slater-type orbitals (STO). In the present work, the one-electron angular momentum states up to $l_{max}$ = 6 and 7 and the orbitals functions with principal quantum number $n$ up to 35 or 40 are used, while in Ref. [19] the authors used $l$ up to 3 and $n_{max}$ up to 11. As for the entanglement in the triplet-spin states in the helium atom, our results are also compared with those in Ref. [17, 18] in which the entanglement was calculated using Kinoshita-type wave functions, and the needed 12-dimensional integrals were treated by Monte-Carlo integration routines. It is seen that the agreement is quite good within those of the uncertainties set in Ref. [17, 18]. Also in Table 8, we show other recent results for comparison [23-26].

In addition to calculations of entropies in He, we have also calculated $S_{vN}$ and $S_L$ for some helium-like ions with Z larger than 2, i.e., Z=3, 4, and 5, etc. Meanwhile, we have noticed that Hofer has reported calculations of entropies for helium-like ions such as the Li$^+$ and Be$^{++}$ ions using Gaussian basis wave functions [26]. Table 9 shows a comparison between our present results and those in Ref. [26]. It is interesting to notice that while our $S_L$ show good agreement with Hofer's results in Ref. [26] for He, Li$^+$ and Be$^{++}$, but the differences for $S_{vN}$ between Hofer's results and ours are quite substantial. As both calculations of $S_{vN}$ are based on the same expression as shown in Eq. (2), such large discrepancies are quite puzzling. It would therefore be desirable to have an independent calculation on $S_{vN}$ for helium-like ions to shed light on the nature of such discrepancy.

Next we have investigated the behavior of entanglement entropies for the abovementioned states when the nuclear charge for the two-electron atom is reduced from Z=2 continuously to the critical value Z=1. In such a scenario, the helium atom that consists of an infinite number of bound states now turns into the hydrogen negative ion, H$^-$ (also called hydride), that has only one bound state, the ground state of H$^-$. All the singly excited states in He (Z=2) now become unbound when Z=1 (see Figure 1). Here in Figures 2 and 3 we show entropy vs 1/Z for the $1s2s$ $^1S$ state and $1s2s$ $^3S$ state, respectively, with parts (a) and (b) for the von Neumann entropy and linear entropy separately. To construct Figures 2 and 3, we have extended calculations for entropies to systems with Z=25, 50, 80 and 100. The results for linear entropy from Z=2 to Z=15 are taken from our earlier investigation [21], and for von Neumann entropy are new calculations. Our detailed numerical results for various Z values for both linear and von Neumann entropies are available upon request.

In an earlier investigation, Osenda and Serra [28] calculated the von Neumann entropy for the $1s2s\ ^3S$ excited state within the $S$-wave model of the two-electron helium-like systems. They showed that when Z is decreased from Z=2 to the critical value Z=1, the von Neumann entropy $S_{vN}$ would approach 1.0. In the present work, we also calculate the entropies for Z=2 continuously to Z=1. In Figures 3(a) and 3(b), we show the $S_L$ and $S_{vN}$ results for the $1s2s\ ^3S$ state. Here, it is seen that at the critical charge Z=1, the values for $S_L$ and $S_{vN}$ are approaching the limiting cases of 0.5 and 1.0 respectively, demonstrating that the system now comprises of an electron bound to the nucleus and the other electron getting free. Such behaviors at the saturated end points are also observed for the singlet-spin $1s2s\ ^1S$ state, as shown in Figures 2(a) and 2(b). Our results are obtained using the full Hamiltonian in the present work, while in Ref. [28] only the $S$-wave component of the potential part in the Hamiltonian was considered. Hence no direct comparison can be made here for the $1s2s\ ^3S$ state. However, it is noted that from Figure 4 in Ref. [28], their $S_{vN}$ value is shown having a value lying below 1.0 at Z=2, while our present $S_{vN}$ value (see Figure 3(a) and Table 2), and those of Refs. [19, 23] all indicate that $S_{vN}$ is larger than 1.0 at Z=2. As the difference between our present work and that of Ref. [28] is their omission of higher order contributions from the multipole expansion for the electron-electron repulsion term (they kept only the monopole term). So an independent investigation including the higher order contributions from the electron-electron repulsion term is called for.

Summarizing our results in Figures 2 and 3, it is concluded that at both limiting cases when 1/Z approaches 0.0 and 1.0, the von Neumann entropy and the linear entropy, respectively, would lead to the saturated values of 1.0 and 0.5 at both end points. The physics for the linear entropy leading to the value of 0.5 when 1/Z approaches zero (Z approaches infinite) was discussed in the earlier work [21]. It is further observed that for the singlet-spin state, there appears a minimum in the plot for entropy vs 1/Z (see Fig. 2), while for the triplet-spin state; a maximum exists in the entropy vs 1/Z plot (see Figure 3). At present, we do not have a definite answer to explain such phenomenon. But we are confident about our numerical results, and it is hoped that our present findings would stimulate further investigations on such phenomenon.

Finally, we discuss the advantages and disadvantages for the present computational approach using $B$-Spline basis sets as compared to other calculations using different wave functions. For accuracy, it seems that for ground state and some lower-lying excited states, the use of Hylleraas functions would lead to the most accurate numerical results [22, 23, 25]. But the use of Hylleraas functions is computationally quite demanding as for a given state in a given ion with charges Z, we need to optimize individually its energy and wave function. On the other hand, the computational scheme used in the present work and in Ref. [21], constructions of two-electron wave functions for several lower-lying states are quite straightforward once the one-electron orthonormal basis set wave functions are established [41]. As was demonstrated in Ref [21], the lowest singlet-spin states $1sns\ ^1S$ (n=1-10) for a given $Z$ were constructed in a single diagonalization of the two-electron Hamiltonian. Similarly, the lowest triplet-spin states $1sns\ ^3S$ (n=2-10) can also be constructed in a single diagonalization of the Hamiltonian. In the present work, we only report our findings for the lowest excited singlet-spin and triplet-spin states for changing $Z$; in response to the investigation of quantum entanglement at critical charge Z for the excited $1s2s\ ^3S$ state in spherical helium (S-wave model for the Hamiltonian [28]). Here, we investigate the counter-part states for various $Z$ values with the full Hamiltonian without approximating the electron-electron interaction operator. Also the present approach using configuration interaction with $B$-Splines basis can be extended to investigate quantum entanglement for higher partial wave excited states ($L > 0$) in a straightforward manner. In the near future, one of our goals is to carry out such an investigation to explore quantum entanglement for high partial wave $(L > 0)$ excited states in natural atoms/ions

## IV. Summary and Conclusion

In the present work, we have reported quantifications of von Neumann entropy and linear entropy for the spatial (electron-electron orbital) entanglement in the ground $1s^2\ ^1S$ state and the $1s2s\ ^{1,3}S$ excited states of the helium atom. The configuration integration (CI) with $B$-Spline basis wave functions are used to represent the ground and excited states of the helium atom. Furthermore, for the excited states, a systematic investigation on von Neumann and linear entropies has been carried out when the nuclear charge is decreased from $Z=2$ continuously to the critical value $Z=1$. It is further shown that when $Z$ is approaching the critical charge, the linear entropies and the von Neumann entropies for these excited states are approaching the limiting values of 0.5 and 1.0, respectively, revealing the system is comprised of an electron bound to the nucleus and the other electron getting free. We believe our present work provides helpful contribution to the current investigations of quantification of quantum entanglement entropies involving indistinguishable particles such as identical fermions or identical bosons. It is also hoped that our present findings would stimulate further investigations from the communities of quantum entanglement in general, and of atomic physics in particular, and from such discussions we might be able to shed light on some interesting development about quantum entanglement in natural atomic systems.

**Acknowledgments**
The present work is supported by the Ministry of Science and Technology in Taiwan.

# APPENDIX A
**Entanglement for the spatial and spin parts of the two electrons in He**

For the two spin-1/2 fermions (electrons) in the helium atom, the wave function $\Phi$ is a product of its spatial (coordinate) part ($\Psi$) and spin part ($\chi$), with $\Phi = \Psi(\vec{r}_1, \vec{r}_2) \chi(\sigma_1, \sigma_2)$, (A.1) where σ symbolizes the spin-up or spin-down components of the single-electron spin wave function. The overall wave function in Eq. (A.1) is anti-symmetric with respect to the interchange of the two electrons. So when the spin wave function is anti-symmetric (the singlet-spin states), the spatial part must be symmetric. Conversely, when the spin wave function is symmetric (the triplet-spin states), the coordinate part of the wave function must be anti-symmetric. The density matrix ($\rho$) is the tensor product of its coordinate part and spin part,

with $\quad\quad\quad \rho = \rho^{(\text{coord})} \otimes \rho^{(\text{spin})}$. (A.2)

Following Refs. [9, 18], entanglement $\xi$ is defined as

$$\xi[|\Phi\rangle] = N\left[S_L(\rho_1) - \left(\frac{N-1}{N}\right)\right],$$ (A.3)

where $N$ is the number of fermions, and $\rho_1$ is the reduced density matrix. Here $N=2$ for the two electrons in the helium atom. Substituting Eq. (2) for $S_L$ to Eq. (A.3), we have

$$\xi[|\Phi\rangle] = 1 - 2\text{Tr}\left[\left(\rho_1^{(\text{coord.})}\right)^2\right]\text{Tr}\left[\left(\rho_1^{(\text{spin})}\right)^2\right]$$ (A.4)

From Refs. [9, 18], the spin parts of the reduced density matrix lead to

$$\text{Tr}\left[\left(\rho_1^{(\text{spin})}\right)^2\right] = 1 \quad \text{(For the triplet-spin states with } S_z= +1 \text{ or } S_z= -1.\text{)};$$ (A.5)

$$\text{Tr}\left[\left(\rho_1^{(\text{spin})}\right)^2\right] = \frac{1}{2} \quad \text{(For the singlet-spin states with } S_z=0 \text{ and for triplet-spin states with } S_z=0),$$ (A.6)

where $S_z$ is the z-component of the total spin $S$ (not to confuse with the entropy $S$) of the two spin-1/2 electrons. Finally, we obtain

$$\xi[|\Phi\rangle] = 1 - 2\text{Tr}\left[\left(\rho_1^{(\text{coord.})}\right)^2\right] \quad \text{(For the triplet-spin states with } S_z= +1 \text{ or } S_z= -1).$$ (A.7)

$$\xi[|\Phi\rangle] = 1 - \text{Tr}\left[\left(\rho_1^{(\text{coord.})}\right)^2\right] \quad \text{(For the singlet-spin states } (S_z=0) \text{ and triplet-spin states with } S_z=0).$$ (A.8)

In the present work, to simplify discussion, we only consider the states with $S_z=0$ (singlet or triplet). For the triplet-spin wave functions with $S_z = \pm 1$, Eq. (A.7) must be used to quantify entanglement. Nevertheless, once $\text{Tr}\left[\left(\rho_1^{(\text{coord.})}\right)^2\right]$ is determined using the same procedure described in the present work, entanglement for triplet-spin states with $S_z = \pm 1$ can trivially be deduced from Eq. (A.7).

## Appendix: B
## Calculations of reduced density matrix for wave functions consist of products of ortho-normal basis

A density matrix is defined as $\rho_{AB} = |\psi\rangle_{AB}\,{}_{AB}\langle\psi|$, with $\langle i|j\rangle = \delta_{ij}$, where |i> and |j> are the members in the one-electron ortho-normal basis set. We can arrange and write the eigenvector $C_{ij}$ for a given state in helium as

$$|\psi\rangle_{AB} = \sum_{ij} C_{ij} \left[|i\rangle_A |j\rangle_B \pm |j\rangle_A |i\rangle_B\right] \tag{B.1}$$

where the + or − signs are for the singlet-spin and triplet-spin states, respectively. The two-component density matrix becomes

$$\begin{aligned}\rho_{AB} &= \sum_{ij} C_{ij}\left[|i\rangle_A |j\rangle_B \pm |j\rangle_A |i\rangle_B\right] \times \sum_{lm} C_{lm}^*\left[{}_A\langle l|\,{}_B\langle m| \pm {}_A\langle m|\,{}_B\langle l|\right] \\ &= \sum_{ijlm} C_{ij} C_{lm}^* \left[|i\rangle_A |j\rangle_B\,{}_A\langle l|\,{}_B\langle m| \pm |i\rangle_A |j\rangle_B\,{}_A\langle m|\,{}_B\langle l| \pm |j\rangle_A |i\rangle_B\,{}_A\langle l|\,{}_B\langle m| + |j\rangle_A |i\rangle_B\,{}_A\langle m|\,{}_B\langle l|\right]\end{aligned} \tag{B.2}$$

Tracing out all the degrees of freedom of particle B,

$$\mathrm{Tr}_B(\rho_{AB}) = \sum_{ilk} C_{ik} C_{lk}^* |i\rangle_A\,{}_A\langle l| \pm \sum_{imk} C_{ik} C_{km}^* |i\rangle_A\,{}_A\langle m| \pm \sum_{jlk} C_{kj} C_{lk}^* |j\rangle_A\,{}_A\langle l| + \sum_{jmk} C_{kj} C_{km}^* |j\rangle_A\,{}_A\langle m| \tag{B.3}$$

and define the reduced density matrix $\rho_A$ as

$$\rho_A = \mathrm{Tr}_B(\rho_{AB}) = \sum_{lmk} (C_{lk} \pm C_{kl})(C_{mk}^* \pm C_{km}^*)|l\rangle\langle m|. \tag{B.4}$$

Here, as particle A and B are identical particles, we hence have $\rho_A = \mathrm{Tr}_B(\rho_{AB}) = \rho_B = \mathrm{Tr}_A(\rho_{AB})$.

Finally, we obtain the matrix elements for the reduced density matrix as

$$(\rho_A)_{ij} = \sum_k (C_{ik} \pm C_{ki})(C_{jk}^* \pm C_{kj}^*) \tag{B.5}$$

Diagonalize $\rho_A$ to find the eigenvalues $\lambda_i$, and the von Neumann entropy can be calculated using Eq. (16) and the linear entropy Eq. (17).

Table 1. Energy levels (in a. u.) for the $1s^2\ ^1S$ ground state and $1s2s\ ^{1,3}S$ excited states of helium

|  | Present | Drake[a] | Dehesa *et. al.*[b] | Manzano *et al.*[c] |
|---|---|---|---|---|
| $1s^2\ ^1S$ | -2.9035820 | -2.903724377 | -2.903724377 | -2.903724377032 |
| $1s2s\ ^1S$ | -2.1459650 | -2.145974046 | -2.145974046 | -2.145974045970 |
| $1s2s\ ^3S$ | -2.1752288 | -2.175229378 | -2.175229378 | -2.175229378225 |

[a] [48]; [b] [18]; [c] [9]

Table 2. Convergence test for linear entropy of $1s^2\ ^1S$ state in helium, cut-off at different $l_{max}$ and $n_{max}$ for hydrogen-like wave functions

| $n_{max}$ | $L_{max}=0$ | $L_{max}=1$ | $L_{max}=2$ | $L_{max}=3$ | $L_{max}=4$ | $L_{max}=5$ | $L_{max}=6$ | $L_{max}=7$ |
|---|---|---|---|---|---|---|---|---|
| 5 | 0.0081787 | 0.0117488 | 0.0161737 | 0.0160678 | | | | |
| 10 | 0.0082581 | 0.0119418 | 0.0161732 | 0.0160678 | 0.0159925 | 0.0159592 | 0.0159438 | 0.0159369 |
| 15 | 0.0082733 | 0.0119808 | 0.0161731 | 0.0160678 | 0.0159925 | 0.0159592 | 0.0159438 | 0.0159369 |
| 20 | 0.0082872 | 0.0120178 | 0.0161730 | 0.0160678 | 0.0159925 | 0.0159592 | 0.0159438 | 0.0159369 |
| 25 | 0.0083635 | 0.0122139 | 0.0161724 | 0.0160678 | 0.0159925 | 0.0159592 | 0.0159438 | 0.0159369 |
| 30 | 0.0088749 | 0.0143232 | 0.0161602 | 0.0160675 | 0.0159925 | 0.0159592 | 0.0159438 | 0.0159369 |
| 35 | 0.0087503 | 0.0162080 | 0.0161075 | 0.0160273 | 0.0159857 | 0.0159588 | 0.0159439 | 0.0159369 |
| 40 | 0.0087420 | 0.0161750 | 0.0160678 | 0.0159925 | 0.0159592 | 0.0159438 | 0.0159369 | |

Final results with estimated uncertainty for $S_L = 0.015937 \pm 0.00004$

Table 3. Convergence test for von Neumann entropy of $1s^2\ ^1S$ state in helium, cut-off at different $l_{max}$ and $n_{max}$ for hydrogen-like wave functions

| $n_{max}$ | $L_{max}=0$ | $L_{max}=1$ | $L_{max}=2$ | $L_{max}=3$ | $L_{max}=4$ | $L_{max}=5$ | $L_{max}=6$ | $L_{max}=7$ |
|---|---|---|---|---|---|---|---|---|
| 5 | 0.0384669 | 0.0605465 | 0.0844488 | 0.0853071 | | | | |
| 10 | 0.0387849 | 0.0616071 | 0.0844518 | 0.0853071 | 0.0851848 | 0.0850829 | 0.0850257 | 0.0849978 |
| 15 | 0.0388455 | 0.0618207 | 0.0844525 | 0.0853071 | 0.0851848 | 0.0850829 | 0.0850257 | 0.0849978 |
| 20 | 0.0389011 | 0.0620226 | 0.0844532 | 0.0853071 | 0.0851848 | 0.0850829 | 0.0850257 | 0.0849978 |
| 25 | 0.0392062 | 0.0630872 | 0.0844565 | 0.0853071 | 0.0851848 | 0.0850829 | 0.0850257 | 0.0849978 |
| 30 | 0.0412993 | 0.0740986 | 0.0845426 | 0.0853064 | 0.0851848 | 0.0850829 | 0.0850257 | 0.0849978 |
| 35 | 0.0413657 | 0.0845338 | 0.0853880 | 0.0852581 | 0.0851623 | 0.0850811 | 0.0850264 | 0.0849977 |
| 40 | 0.0413388 | 0.0844416 | 0.0853071 | 0.0851848 | 0.0850829 | 0.0850257 | 0.0849978 | |

Final result with estimated uncertainty for $S_{vN} = 0.084998 \pm 0.0001$

Table 4. Convergence test for linear entropy of $1s2s\ ^1S$ state in helium, cut-off at different $l_{max}$ and $n_{max}$ for hydrogen-like wave functions

| $n_{max}$ | $L_{max}=0$ | $L_{max}=1$ | $L_{max}=2$ | $L_{max}=3$ | $L_{max}=4$ | $L_{max}=5$ | $L_{max}=6$ | $L_{max}=7$ |
|---|---|---|---|---|---|---|---|---|
| 5  | 0.4869008 | 0.4875597 | 0.4886460 | 0.4887119 |           |           |           |           |
| 10 | 0.4867619 | 0.4875802 | 0.4886461 | 0.4887121 | 0.4887283 | 0.4887339 | 0.4887362 | 0.4887370 |
| 15 | 0.4867410 | 0.4875846 | 0.4886461 | 0.4887121 | 0.4887283 | 0.4887339 | 0.4887362 | 0.4887370 |
| 20 | 0.4867233 | 0.4875887 | 0.4886460 | 0.4887121 | 0.4887283 | 0.4887339 | 0.4887362 | 0.4887370 |
| 25 | 0.4866474 | 0.4876114 | 0.4886458 | 0.4887123 | 0.4887283 | 0.4887339 | 0.4887362 | 0.4887370 |
| 30 | 0.4869578 | 0.4879393 | 0.4886340 | 0.4887133 | 0.4887285 | 0.4887339 | 0.4887362 | 0.4887370 |
| 35 | 0.4875407 | 0.4886226 | 0.4886904 | 0.4887125 | 0.4887271 | 0.4887339 | 0.4887363 | 0.4887370 |
| 40 | 0.4875435 | 0.4886373 | 0.4887118 | 0.4887283 | 0.4887339 | 0.4887362 | 0.4887370 |           |

Final result with estimated uncertainty for $S_L = 0.488737 \pm 0.00001$

Table 5. Convergence test for von Neumann entropy of $1s2s\ ^1S$ state in helium, cut-off at different $l_{max}$ and $n_{max}$ for hydrogen-like wave functions

| $n_{max}$ | $L_{max}=0$ | $L_{max}=1$ | $L_{max}=2$ | $L_{max}=3$ | $L_{max}=4$ | $L_{max}=5$ | $L_{max}=6$ | $L_{max}=7$ |
|---|---|---|---|---|---|---|---|---|
| 5  | 0.9819964 | 0.9872351 | 0.9916100 | 0.9918722 |           |           |           |           |
| 10 | 0.9818511 | 0.9873554 | 0.9916201 | 0.9918725 | 0.9919094 | 0.9919161 | 0.9919173 | 0.9919171 |
| 15 | 0.9818318 | 0.9873802 | 0.9916220 | 0.9918726 | 0.9919094 | 0.9919161 | 0.9919173 | 0.9919171 |
| 20 | 0.9818163 | 0.9874039 | 0.9916239 | 0.9918727 | 0.9919094 | 0.9919161 | 0.9919173 | 0.9919171 |
| 25 | 0.9817638 | 0.9875316 | 0.9916313 | 0.9918730 | 0.9919095 | 0.9919161 | 0.9919173 | 0.9919171 |
| 30 | 0.9827186 | 0.9891170 | 0.9916731 | 0.9918773 | 0.9919097 | 0.9919161 | 0.9919173 | 0.9919171 |
| 35 | 0.9839065 | 0.9915202 | 0.9918324 | 0.9918838 | 0.9919069 | 0.9919149 | 0.9919173 | 0.9919171 |
| 40 | 0.9839099 | 0.9915461 | 0.9918718 | 0.9919094 | 0.9919161 | 0.9919173 | 0.9919171 |           |

Final result with estimated uncertainty for $S_{vN} = 0.991917 \pm 0.00001$

Table 6. Convergence test for linear entropy of $1s2s\ ^3S$ state in helium, cut-off at different $l_{max}$ and $n_{max}$ for hydrogen-like wave functions

| $n_{max}$ | $L_{max}=0$ | $L_{max}=1$ | $L_{max}=2$ | $L_{max}=3$ | $L_{max}=4$ | $L_{max}=5$ | $L_{max}=6$ | $L_{max}=7$ |
|---|---|---|---|---|---|---|---|---|
| 5 | 0.5000000 | 0.5001432 | 0.5003710 | 0.5003761 | | | | |
| 10 | 0.5000001 | 0.5001612 | 0.5003711 | 0.5003761 | 0.5003761 | 0.5003760 | 0.5003760 | 0.5003760 |
| 15 | 0.5000001 | 0.5001647 | 0.5003711 | 0.5003761 | 0.5003761 | 0.5003760 | 0.5003760 | 0.5003760 |
| 20 | 0.5000001 | 0.5001680 | 0.5003711 | 0.5003761 | 0.5003761 | 0.5003760 | 0.5003760 | 0.5003760 |
| 25 | 0.5000001 | 0.5001849 | 0.5003711 | 0.5003761 | 0.5003761 | 0.5003760 | 0.5003760 | 0.5003760 |
| 30 | 0.5000012 | 0.5003172 | 0.5003724 | 0.5003761 | 0.5003761 | 0.5003760 | 0.5003760 | 0.5003760 |
| 35 | 0.5000030 | 0.5003710 | 0.5003761 | 0.5003761 | 0.5003760 | 0.5003760 | 0.5003760 | 0.5003760 |
| 40 | 0.5000030 | 0.5003710 | 0.5003761 | 0.5003761 | 0.5003760 | 0.5003760 | 0.5003760 | |

Final result with estimated uncertainty for $S_L = 0.500376 \pm 0.00001$

Table 7. Convergence test for von Neumann entropy of $1s2s\ ^3S$ state in helium, cut-off at different $l_{max}$ and $n_{max}$ for hydrogen-like wave functions

| $n_{max}$ | $L_{max}=0$ | $L_{max}=1$ | $L_{max}=2$ | $L_{max}=3$ | $L_{max}=4$ | $L_{max}=5$ | $L_{max}=6$ | $L_{max}=7$ |
|---|---|---|---|---|---|---|---|---|
| 5 | 1.0000010 | 1.0022783 | 1.0053749 | 1.0055174 | | | | |
| 10 | 1.0000017 | 1.0025355 | 1.0053763 | 1.0055174 | 1.0055267 | 1.0055273 | 1.0055272 | 1.0055272 |
| 15 | 1.0000019 | 1.0025854 | 1.0053767 | 1.0055174 | 1.0055267 | 1.0055273 | 1.0055272 | 1.0055272 |
| 20 | 1.0000020 | 1.0026320 | 1.0053770 | 1.0055174 | 1.0055267 | 1.0055273 | 1.0055272 | 1.0055272 |
| 25 | 1.0000030 | 1.0028693 | 1.0053787 | 1.0055174 | 1.0055267 | 1.0055273 | 1.0055272 | 1.0055272 |
| 30 | 1.0000258 | 1.0046643 | 1.0054179 | 1.0055177 | 1.0055267 | 1.0055273 | 1.0055272 | 1.0055272 |
| 35 | 1.0000598 | 1.0053732 | 1.0055171 | 1.0055263 | 1.0055272 | 1.0055273 | 1.0055272 | 1.0055272 |
| 40 | 1.0000599 | 1.0053732 | 1.0055174 | 1.0055267 | 1.0055273 | 1.0055272 | 1.0055272 | |

Final result with estimated uncertainty for $S_{vN} = 1.005527 \pm 0.00001$

Table 8. Comparison for linear entropy $S_L$ and von Neumann entropy $S_{vN}$

| | $S_L$ | $S_{vN}$ |
|---|---|---|
| $1s^2\ ^1S$ | $0.015937 \pm 0.00004$[a] | $0.084998 \pm 0.0001$[a] |
| | $0.015943 \pm 0.00004$[b] | |
| | $0.0159156 \pm 0.0000010$[c] | |
| | $0.015914 \pm 0.000044$[d] | |
| | $0.01606$[e] | $0.0785$[e] |
| | $0.01591564$[f] | $0.08489987$[f] |
| | $0.0159172$[g] | |
| | $0.0159157$[h] | $0.0848999$[h] |
| | $0.01595052$[i] | $0.06749889$[i] |
| | | $0.0675$[j] |
| $1s2s\ ^1S$ | $0.488737 \pm 0.00001$[a] | $0.991917 \pm 0.00001$[a] |
| | $0.488736$[b] | |
| | $0.48866 \pm 0.00030$[d] | |
| | $0.48871$[e] | $0.991099$[e] |
| | $0.48874040$[f] | $0.99191721$[f] |
| $1s2s\ ^3S$ | $0.500376 \pm 0.00001$[a] | $1.005527 \pm 0.00001$[a] |
| | $0.50038 \pm 0.00015$[d] | |
| | $0.500378$[e] | $1.00494$[e] |
| | $0.50037593$[f] | $1.00552680$[f] |

[a] Present calculations using Eq. (16) for $S_{vN}$ or Eq. (17) for $S_L$
[b] Lin *et. al.* [21], using CI basis with B-spline and Eqs. (3) and (7).
[c] Lin *et. al.* [22], using Hylleraas functions and Eqs. (3) and (7)
[d] Dehesa *et. al.*[18] using Kinoshita-type wave functions and Eq. (3) and (7)
[e] Benenti *et. al.*[19] using CI basis with STO and Eq. (16) or Eq. (17)
[f] Lin and Ho [23] using Hylleraas functions and Schmidt-Slater decomposition method
[g] Kościk [24] using Hylleraas functions and Schmidt-Slater decomposition method
[h] Koscik and Okopinska [25] using Hylleraas functions and Schmidt-Slater decomposition method
[i] Hofer [26] using Gaussian basis
[j] Huang *et. al.* [20] using Gaussian basis and Eq. (16)

Table 9. Energy, von Neumann entropy and linear entropy of He, Li$^+$, B$^{2+}$ and B$^{3+}$

|  |  | $1s^2\ ^1S$ |  | $1s2s\ ^1S$ | $1s2s\ ^3S$ |
|---|---|---|---|---|---|
|  |  | Present | Other | Present | Present |
| He | E(a.u.) | -2.90361799 | -2.90361147[a] | -2.1459650 | -2.1752288 |
|  | $S_L$ | 0.015937 | 0.01590025[a] | 0.4887362 | 0.5003760 |
|  | $S_{vN}$ | 0.084998 | 0.06732135[a] | 0.9919173 | 1.0055272 |
| Li$^+$ | E(a.u.) | -7.27974065 | -7.27933199[a] | -5.040859 | -5.11073 |
|  | $S_L$ | 0.006547 | 0.00655259[a] | 0.493031 | 0.500221 |
|  | $S_{vN}$ | 0.039532 | 0.03184967[a] | 0.997094 | 1.003424 |
| Be$^{2+}$ | E(a.u.) | -13.65534197 | -13.65492843[a] | -9.184847 | -9.29716 |
|  | $S_L$ | 0.003561 | 0.00356247[a] | 0.495638 | 0.500138 |
|  | $S_{vN}$ | 0.023162 | 0.01886441[a] | 0.999177 | 1.002238 |
| B$^{3+}$ | E(a.u.) | -22.03071007 |  | -14.578493 | -14.73389 |
|  | $S_L$ | 0.002237 |  | 0.497071 | 0.500094 |
|  | $S_{vN}$ | 0.015331 |  | 0.999975 | 1.001569 |

[a]Hofer [26] using Gaussian basis

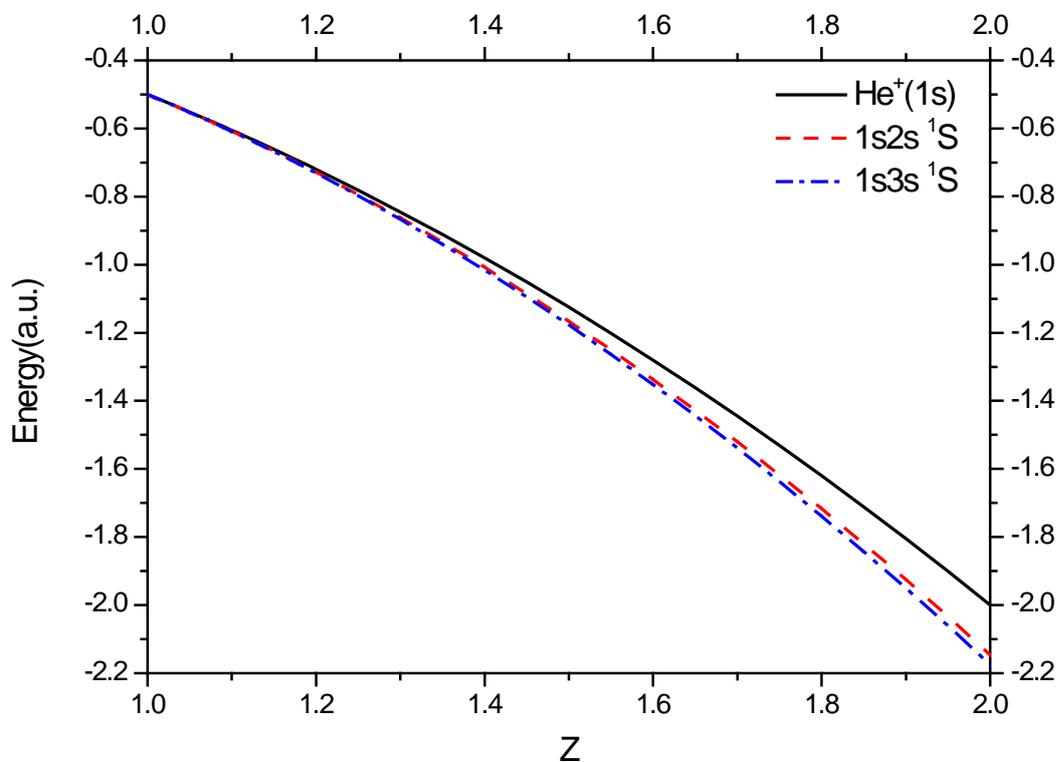

Figure 1. Energy levels for the $1s2s$ and $1s3s\ ^{1,3}S$ states from $Z=2$ to $Z=1$.

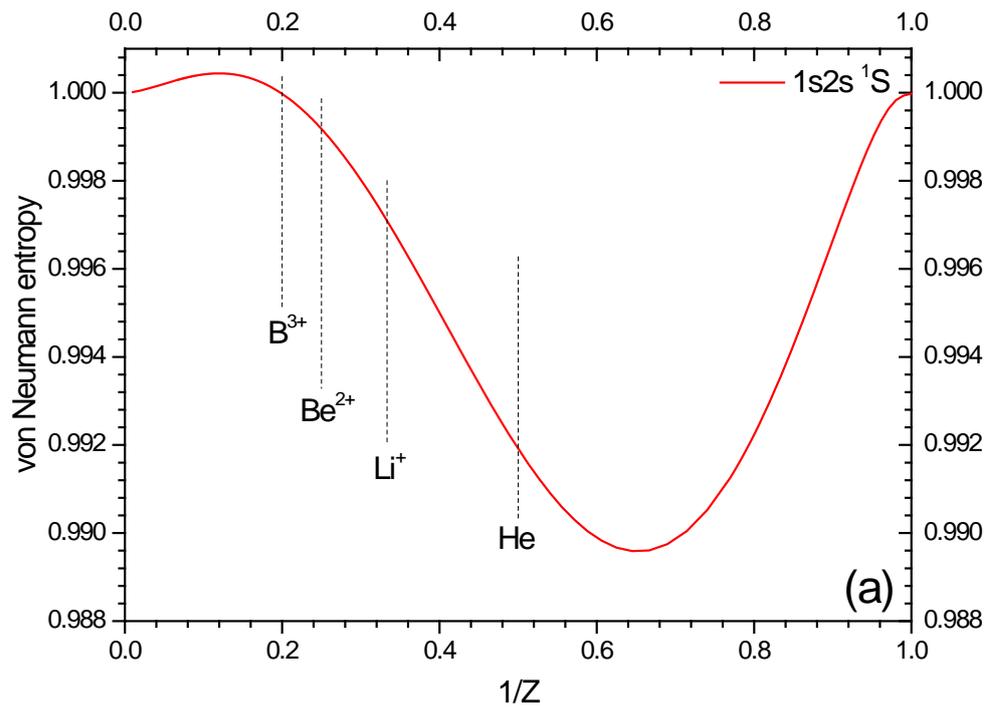

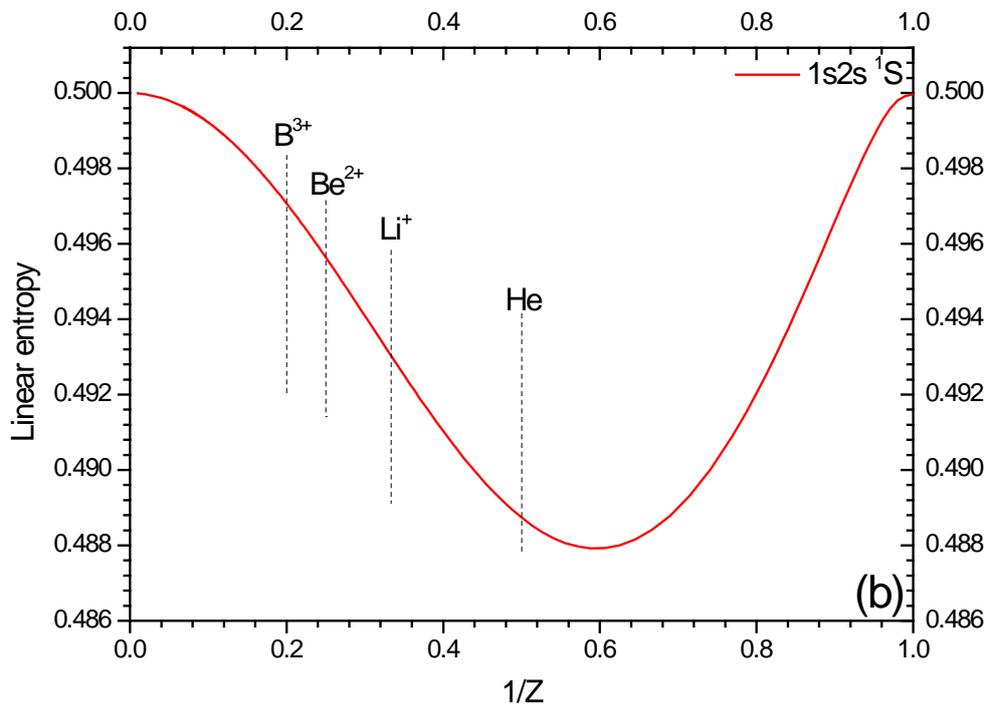

Fig. 2. (a) von Neumann entropy and (b) linear entropy *vs* 1/Z for the $1s2s$ $^1S$ state of helium. The dashed lines show the entropy values for the He atom, $Li^+$, $Be^{2+}$ and $B^{3+}$ ions.

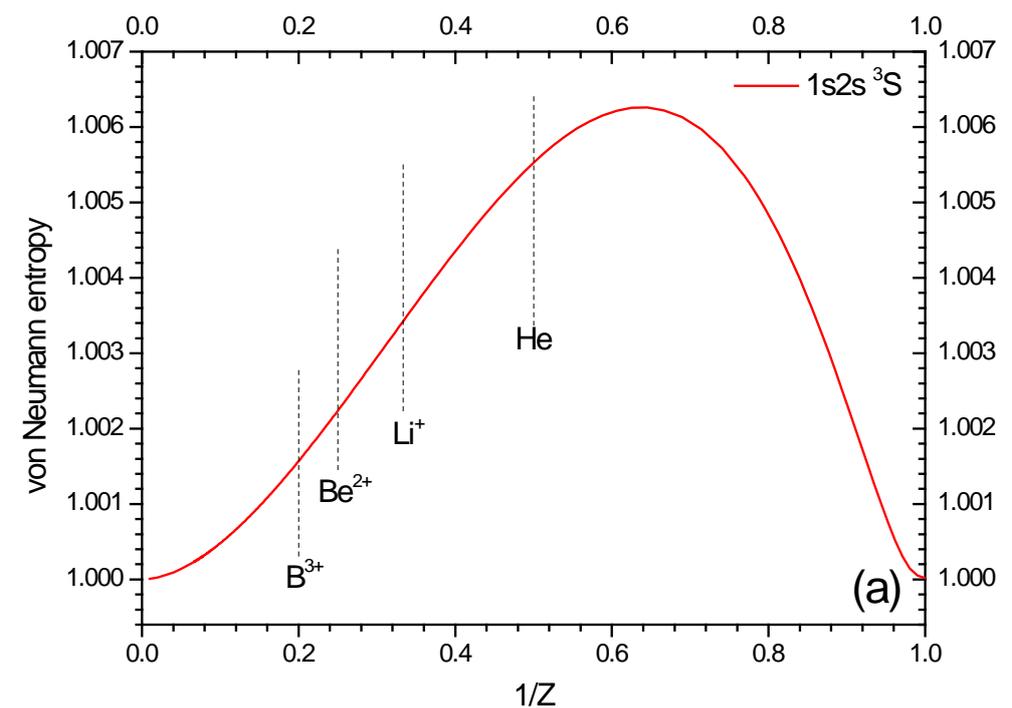

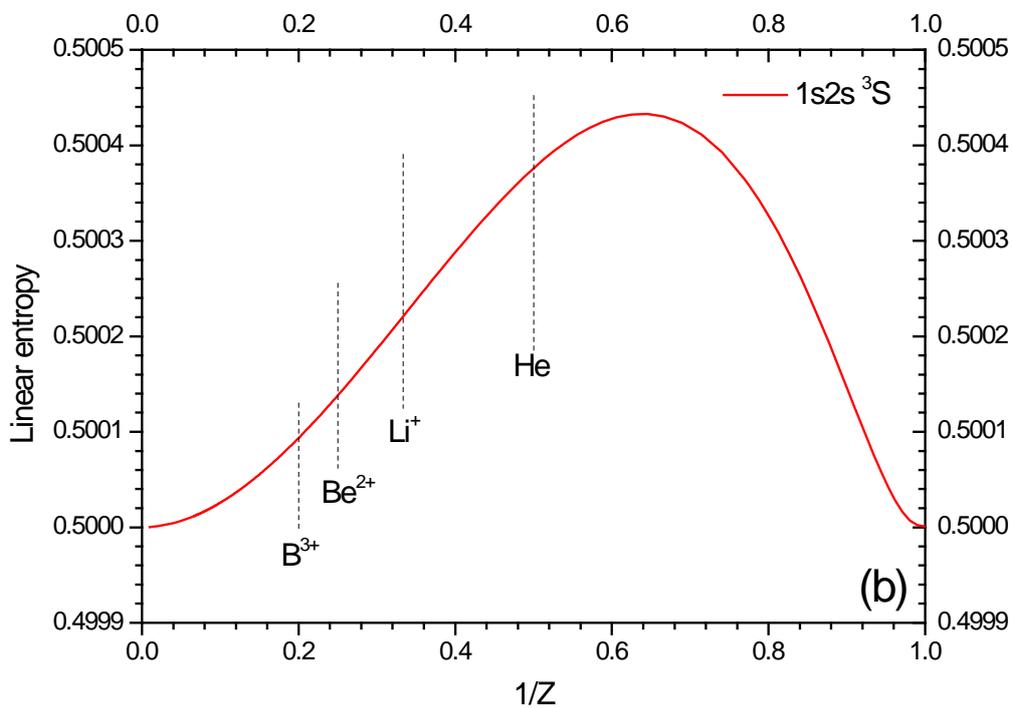

Fig. 3. (a) von Neumann entropy and (b) linear entropy *vs* 1/Z for the $1s2s\ ^3S$ state of helium. The dashed lines show the entropy values for the He atom, $Li^+$, $Be^{2+}$ and $B^{3+}$ ions